\documentstyle[12pt,epsf]{article}
 \catcode`@=11
\long\def\@caption#1[#2]#3{\par\addcontentsline{\csname
  ext@#1\endcsname}{#1}{\protect\numberline{\csname
  the#1\endcsname}{\ignorespaces #2}}\begingroup
    \small
    \@parboxrestore
    \@makecaption{\csname fnum@#1\endcsname}{\ignorespaces #3}\par
  \endgroup}
\catcode`@=12
\newcommand{\bi}{\bibitem}
\begin{document}
\setlength{\baselineskip}{0.27in}

\newcommand{\beq}{\begin{equation}}
\newcommand{\eeq}{\end{equation}}
\newcommand{\beqa}{\begin{eqnarray}}
\newcommand{\eeqa}{\end{eqnarray}}
\newcommand{\lsim}{\begin{array}{c}\,\sim\vspace{-21pt}\\<
\end{array}}
\newcommand{\gsim}{\begin{array}{c}\sim\vspace{-21pt}\\>
\end{array}}

\begin{titlepage}
{\hbox to\hsize {\hfill UCSD-97-02}}
{\hbox to\hsize {\hfill CALT-68-2095}}
\begin{center}
\vglue .06in
{\Large \bf The Octet Structure Function and Radiative Quarkonia Decays}
\\[.5in]

\begin{tabular}{c}
{\bf Ira Z. Rothstein}\\[.05in]
{\it Department of Physics}\\
{\it University of California at San Diego}\\
{\it La Jolla, CA 92122 }\\[.15in]
\end{tabular}
\vskip 0.25cm

\begin{tabular}{c}
{\bf Mark B. Wise}\\[.05in]
{\it California Institute of Technology}\\
{\it Pasadena, CA 91125 }\\[.15in]
\end{tabular}
  \vskip 0.25cm

{\bf Abstract}\\[-0.05in]

\begin{quote}
The Operator Product Expansion, in conjunction with the power counting
of non-relativistic field theory, is used to examine the end-point
region of the radiative decay of heavy quark bound states with $^3S_1$
quantum numbers, $Q\bar{Q}\rightarrow \gamma+X$. 
We identify an infinite class of operators that determine the shape
of the photonic end-point spectrum. These operators can be resummed to
form an octet structure function which parameterizes the
energy of the dynamical gluon content of the leading octet
Fock state component of the quarkonium.
This color-octet contribution is important when 
the photon spectrum is examined with a resolution given by
$\Delta E_\gamma \sim m_Qv^2$, where $v^2$ is the relative quark velocity
squared.
The formalism used makes
explicit the shift of the end-point from its partonic to
its hadronic value.
\end{quote}
\end{center}
\end{titlepage}
\newpage

The direct photon spectrum in  $\Upsilon (1S)$ decay has been studied 
experimentally\cite{nem} and compared with predictions based upon the
color-singlet model\cite{bro,phot,cat,field}. These predictions were refined by the
inclusion of hadronization effects \cite{field} which one would
expect to be important in the end-point region. In this letter
we discuss an effect which is formally larger near the end-point which has
been heretofore neglected. This effect
stems from higher Fock states in the heavy quark bound state.
Using an operator product expansion (OPE) in conjunction with 
non-relativistic QCD (NRQCD) \cite{bbl}, we show that the end-point
spectrum receives a leading order contribution from a color
octet structure function. Furthermore, the formalism shows explicitly
the shifting of the maximal photon energy from its partonic to its
hadronic value. Thus, studying the endpoint region of the spectrum
becomes a useful laboratory for studying non-perturbative QCD effects.

Direct photons in quarkonia decay arise from the electromagnetic
coupling of both the heavy and light quarks. The spectrum stemming from
the couplings to the light quarks has been previously studied in 
\cite{cat} using the leading logarithmic approximation. 
This fragmentation contribution to the photon energy
spectrum is suppressed relative
to that coming from the coupling to heavy quarks, especially near
the photonic endpoint where the hadronic invariant mass is
highly restricted. In this letter, we will focus on the spectrum due to the
coupling to the heavy quarks.

The photonic decay of quarkonia has the merit that the decay rate contains
an external variable, namely the photon energy, which allows one
to analytically continue to the unphysical regime where the 
OPE is trustworthy. This is tantamount to the statement
that we may smear over a range of hadronic invariant mass, 
resulting in an average which
is dual to the parton model \cite{bg}. This is in contrast to the
total hadronic decay calculations  which assume duality on a point by point
basis.

For the case at hand, we begin by considering the imaginary
part of the time ordered product of two electromagnetic
currents
\beq
T=i\int d^4x e^{-iq\cdot x}
\langle \Upsilon \mid T(J^\mu(x)J_\mu(0)) \mid \Upsilon 
\rangle,
\eeq
where $J^\mu=\bar{b}\gamma_\mu b$ and the normalization convention for
 upsilon states is $\langle \Upsilon(p)\mid \Upsilon(p^{\prime})\rangle
=(2\pi)^3 \delta^3(\vec{p}^{\prime}-\vec p)$. In eq.(1) and hereafter, 
spin averaging
of the $\Upsilon$ matrix element is understood. At fixed, $q^2=0$,
$T$ has two cuts which span the entire real $q_0$ axis (we work 
in the rest frame of the $\Upsilon$) in the complex $q_0$ plane.
There is a cut resulting from the physical process $\Upsilon \rightarrow
\gamma +X$ in the region $-m_\Upsilon /2\leq q_0 \leq m_\Upsilon /2$
(neglecting the pion mass) 
and another from the process $\gamma+\Upsilon \rightarrow X$
for which the final state, $X$, has invariant
 mass larger than $m_\Upsilon$. This overlap
obstructs the usual dispersion relation analysis since the
discontinuity across the cut  in the physical region is polluted by the
other process. However, this obstruction is a red herring.
To see this, we note that the $\Upsilon\rightarrow \gamma+X$ decay rate 
can be extracted from the
time ordered product
\beq 
\label{Tp}
T^\prime= 
i\int d^4x e^{-iq\cdot x}
\langle \Upsilon_{12} \mid T(J^\mu(x)_{12}J_\mu(0)_{34}) \mid \Upsilon_{34}
\rangle,
\eeq
where $J^\mu_{ij}=\bar{b}_{i}\gamma_\mu b_j$ and 
$b_i$ are differing degenerate b-type quark species.
$T'$ has two cuts, however they are now separated as shown in Figure 1. 
To get the  $\Upsilon\rightarrow \gamma+X$ rate we simply perform the 
contour integral over a contour which only picks up a contribution from
the part we are interested in. 
We may then relate the decay
rate to the imaginary part of the $T^\prime$ via
\beq 
\label{rate}
\frac{d\Gamma}{dE_\gamma}=\frac {2 e_b^2 \alpha}{\pi}E_{\gamma} Im T^\prime,
\eeq  
where $e_b=-1/3$ is the electric charge of the $b$-quark and $\alpha$ is the
fine structure constant. 

\begin{figure}[t]
   \vspace{0cm}
   \epsfysize=10cm
   \epsfxsize=10cm
   \centerline{\epsffile{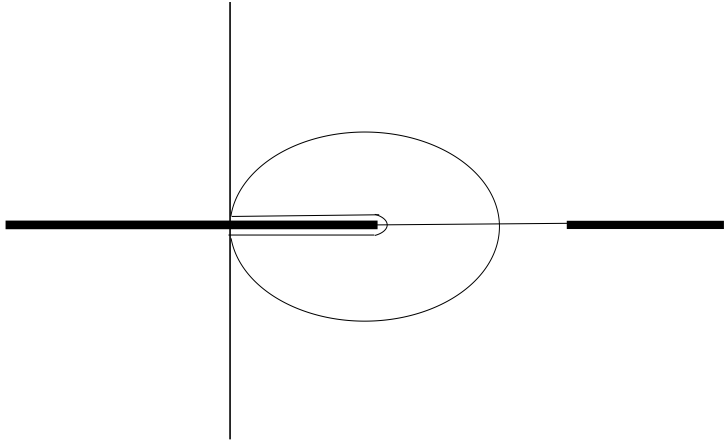}}
   \vspace*{0cm}
\caption{\label{diagplot}Analytic structure of $T^\prime$ in the complex 
$q_0$ plane and the 
appropriate choice of contour. The cut on the left hand side begins at
$q_0=m_\Upsilon /2$ whereas the cut on the right hand side begins at
$q_0=3m_\Upsilon /2$.}
\end{figure}

We now may perform an OPE on the product of currents and calculate
the Wilson coefficients using perturbative QCD. The OPE is valid in the
unphysical region which is related to the physical process via
the contour shown in Figure 1. The contour approaches the cut
at one point, and thus the level of rigor here is not at
the same level as in deep inelastic scattering but should
be considered at the same level as the predictions for
semi-leptonic heavy meson decay \cite{chay,bigi}. Performing the contour
integral over the photon energy corresponds to the smearing
mentioned above. We may choose different weightings
for the contour to extract information regarding the spectrum.
Furthermore, as will be seen below, the convergence of the OPE
will be dictated by the size of the smearing region.

For a given photon energy $E_\gamma$, the final hadronic
invariant mass is given by $m_X^2=m^2_\Upsilon (1-2E_{\gamma}/
{m_\Upsilon})$. Consequently, the end-point region of the
photon energy spectrum corresponds to low invariant mass hadronic final states.
In this region, the photon energy spectrum eq.(3) must be smeared
over a range $\Delta E_\gamma$. We will see that the OPE 
gives us quantitative information regarding the necessary size of 
the smearing region. Indeed, the convergence of the OPE is
predicated upon the prudent choice of smearing functions.

The leading order diagrams come from those shown in Figure 2b. This
decay mode is dominated by the color-singlet Fock state. 
NRQCD velocity scaling rules place this contribution to
the decay rate of order, $\alpha \alpha_s^2 v^3/\pi^2$. Away from the
 end-point this is the leading piece. However, near
the end point there will also be a contribution stemming from
those diagrams shown in Figure 2a which are dominated by the
Fock state where the quarks are in a relative octet state.
This color-octet process yields contributions to the decay rate of order 
$\alpha \alpha_sv^6/ \pi$ and $\alpha \alpha_sv^7/ \pi$ for the
$^1S_0$ and $^3P_J$ intermediate states, respectively \cite{bbl}. 
If we use $v^2\simeq \alpha_s/\pi$,
then the contributions are  down by a factors of $v$ and $v^2$ relative
to the color-singlet process. However, this naive counting does not hold
for the end-point region of the photon energy spectrum
 because the color octet contribution is highly
singular there. Indeed, as will be shown below, after smearing
over a region of photon energies, $\Delta E_\gamma \sim m_bv^2$, the octet
process is of the same order as the singlet in the endpoint region.

The leading order octet contribution to $T^\prime$ is given by
\beqa
\label{lead8}
 T^\prime_8=4g_s^2 \langle \Upsilon_{12} \mid  \left[ \bar{b}_1 
\gamma^\alpha \frac{(p\!\!\! /+iD\!\!\! /-q\!\!\! /+m_b)}
{(p+iD-q)^2-m_b^2}T^A 
\gamma_\nu b_2\right]&&  \\ \nonumber   \frac{g^{\mu \nu}}{(2p-q+iD)^2+i\epsilon}
&&
\left[\bar{b}_3 T^A \gamma_\mu \frac{(p\!\!\! /
  +iD\!\!\! /-q\!\!\! /+m_b)}{(p
+iD-q)^2-m_b^2}
 \gamma_\alpha b_4 \right]\mid \Upsilon_{34}
\rangle
,
\eeqa
where each of the heavy quarks carries momentum $p=(m_b,\vec 0)$
and $D$ denotes a covariant derivative.
$T^\prime_8$ receives contributions from $^1S_0$ and $^3P_J$ intermediate
states so we write 
\beq
T^\prime_8=T_8^\prime(^3P_J)+T_8^\prime(^1S_0).
\eeq

 We now match onto the 
NRQCD by expanding the fields $b$ in terms of non-relativistic
fields $\psi$ and $\chi$ for the particle and anti-particle
respectively. In making the transition to the effective theory, 
we keep only the Dirac structures relevant to the  decay at hand. 
The relation between
b-quark fields in full QCD and NRQCD produces additional 
factors of $D$. After imposing spin symmetry
we find that the contribution to the imaginary part of $T^\prime$ 
from the octets Fock state are given by
\beqa
\label{Imlead}
Im T^\prime_8(^1S_0)= g_s^2 C_{^1S_0} \int dk^+_8 
 O(^1S_0) f(k^+_8)_{^1S_0} 
\delta( m_b
- E_\gamma+k^+_8/2),
\eeqa
\beqa
Im T^\prime_8(^3P_J)= g_s^2  C_{^3P_J} \int dk^+_8  O(^3P_J) 
f(k^+_8)_{^3P_J}  
\delta( m_b
- E_\gamma+k^+_8/2),
\eeqa
where we have defined
\beq
f(k^+_8)_{^3 P_J}= \langle \Upsilon \mid \left[ \psi^\dagger \sigma_i \frac{i}{2}
\stackrel{\leftrightarrow}{D}_j T^A 
\chi \right] \delta(k^+_8- n\cdot iD)
\left[\chi^\dagger  \sigma_i\frac{i}{2} \stackrel{\leftrightarrow}{D}_j T^A 
\psi \right]\mid \Upsilon \rangle/O(^3P_J),
\eeq
\beq
 O(^3P_J)= \langle \Upsilon \mid \left[ \psi^\dagger \sigma_i\frac{i}{2}
 \stackrel{\leftrightarrow}{D}_j T^A 
\chi \right]
\left[\chi^\dagger  \sigma_i \frac{i}{2} \stackrel{\leftrightarrow}{D}_j T^A 
\psi \right]\mid \Upsilon 
\rangle,
\eeq
\beq
f(k^+_8)_{^1 S_0}= \langle \Upsilon \mid \left[ \psi^\dagger 
T^A 
\chi \right] \delta(k^+_8- n\cdot iD)
\left[\chi^\dagger  T^A 
\psi \right]\mid \Upsilon \rangle/O(^1S_0),
\eeq
\beq
 O(^1S_0)= \langle \Upsilon \mid \left[ \psi^\dagger T^A 
\chi \right]
\left[\chi^\dagger  T^A 
\psi \right]\mid \Upsilon 
\rangle,
\eeq
and the light-like four-vector $n=(2p-q)/m_b$ 
is taken to be, $n= (1,0,0,1)$. The four-vector, $k^+_8$, is the light cone momentum of the dynamical gluon
in the octet Fock state. Furthermore we
have kept only the leading twist operators, which is to say we have
dropped terms of order $(k^+_8)^2$ in the delta function. The constants
$C_{^3P_J}$ and $C_{^1S_0}$ have the values
\beq
\label{pathetic}
C_{^3P_J}=\frac{7 \pi}{18 m_b^5},~~C_{^1S_0}=\frac{\pi}{2m_b^3}.
\eeq

The functions $f(k^+_8)_{^3P_J}$ and  $f(k^+_8)_{^1S_0}$ are the normalized probabilities that
the dynamical gluon in the octet Fock state  has light-cone 
momentum  $k^+_8$. A strong analogy may be made
with the deep-inelastic structure functions by studying the Fourier
transform of the $f(k^+_8)$'s. In the gauge, $n\cdot A=0$, 
the Fourier transform introduces a factor of $exp(tn\cdot\partial)$.
This translates the combination of quark fields $[\chi^{\dagger}\psi]$
from the origin to the spacetime point $tn$ giving, for example in the
case of $^3P_J$,
\beq
\hat{f}(t)_{^3 P_J}=\langle \Upsilon \mid 
\left[ \psi^\dagger (0) \sigma_i \frac{i}{2}\stackrel{\leftrightarrow}{
D}_j T^A 
\chi(0) \right]\frac{P exp[-ig_s\int^t_0dt^\prime n\cdot A(t^{\prime}n )]}
{O(^3P_J)}
\left[\chi^\dagger(tn)  \sigma_i \frac{i}{2}\stackrel{\leftrightarrow}{
D}_j T^A 
\psi(tn) \right]\mid \Upsilon 
\rangle.
\eeq
In eq.(13) a path-ordered exponential has been inserted to restore
gauge invaiance.

\begin{figure}[t]
   \vspace{0cm}
   \epsfysize=10cm
   \epsfxsize=10cm
   \centerline{\epsffile{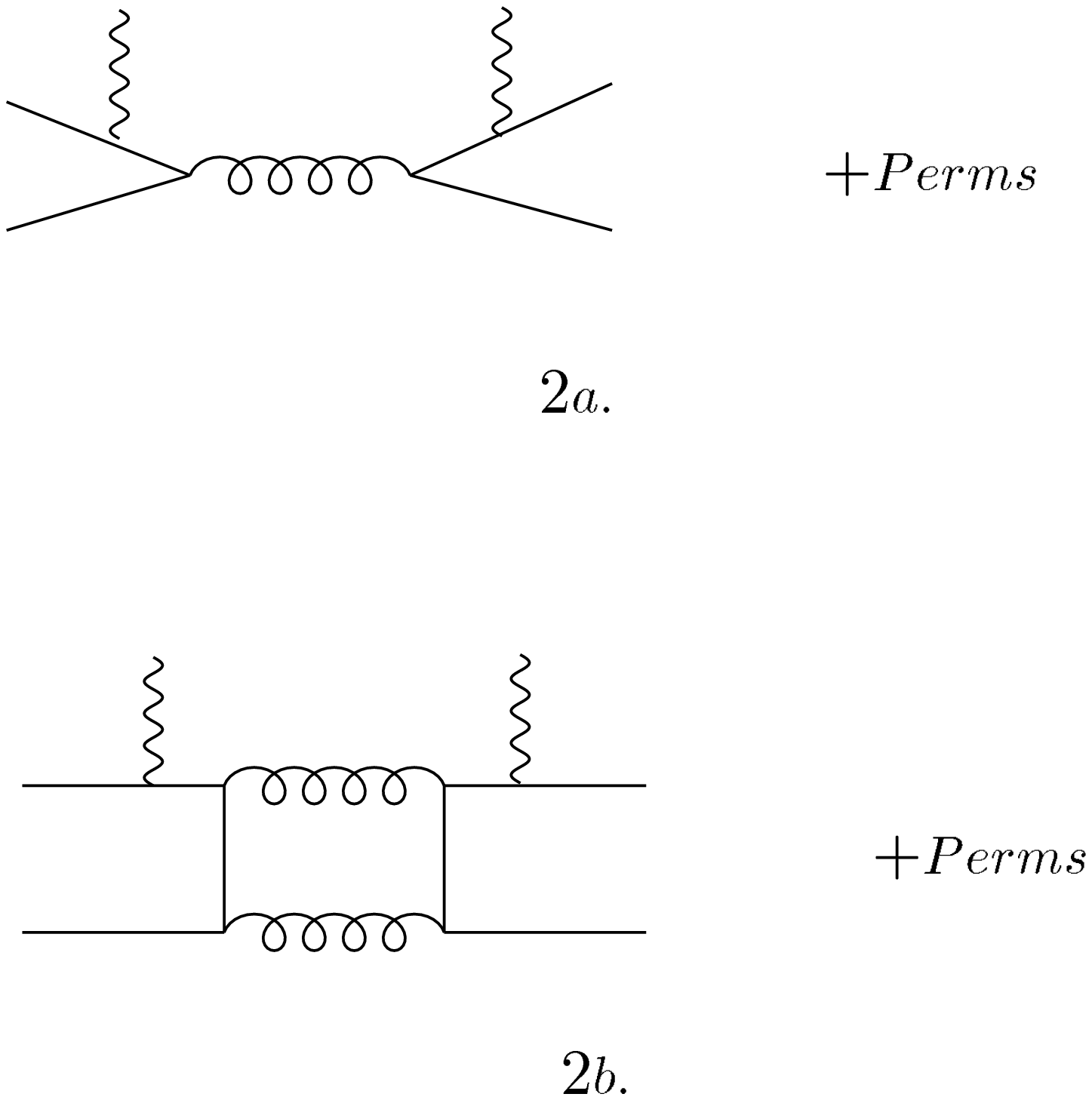}}
   \vspace*{0cm}
\caption{\label{diagplot}2a) Leading order diagram for the octet 
contribution. 2b) Leading order Diagram for the singlet contribution.}
\end{figure}

The analysis of the color singlet contribution is simplified
by the fact that in this case we may use the vacuum saturation
approximation, which is a well controlled expansion in
$v^2$ for this case \cite{bbl}. The net result of summing the
leading twist corrections is simply to shift the 
maximal photon energy from its partonic to hadronic value.
To see this, we note that the imaginary part of Figure 2a
can be written as 
\beq
\label{Imsing}
ImT_1^\prime (^3S_1)= g_s^4 G(E_\gamma)
\sum_{n=0}^\infty \frac{(-1)^n}{n!}
\frac{\partial^n}{\partial(2 E_\gamma)^n}
\theta(2m_b-2E_\gamma ) 
\langle \Upsilon \mid \left[ \psi^\dagger \sigma_i  
\chi \right]
(n\cdot i\partial)^n  
\left[\chi^\dagger  \sigma_i   
\psi \right]\mid \Upsilon 
\rangle,
\eeq
where the leading order Wilson coefficient, $G(E_\gamma)$,
is a smooth function
of $E_\gamma$, that does not vanish at the end-point.
After factorizing this matrix element only the time derivatives contribute
(for $\Upsilon$ states at rest) giving factors of the binding energy
\cite{grem}.
Consequently eq.(14) becomes
\beq
ImT_1^\prime(^3S_1)= g_s^4 G(E_\gamma)
\theta(m_\Upsilon-2E_\gamma) \langle \Upsilon \mid \psi^\dagger \sigma_i 
 \chi
\mid 0 \rangle\langle 0 \mid \chi^\dagger \sigma_i  \psi
\mid \Upsilon \rangle,
\eeq
showing that the whole effect of the leading twist contributions
for the color singlet part of the amplitude is to shift the end-point
from its partonic value, $m_b$, to the physical one $m_{\Upsilon}/2$.

We are now in a position to consider the relative sizes of
$T_1^\prime(^3S_1)$, $T_8^\prime (^3P_J)$ and $T_8^\prime (^1S_0)$ 
in the end-point region.
If the smearing region is large enough, then we would expect that
the singular nature of the octet contribution will not compensate for
naive $v$ suppression of the octet state contributions. 
Too small of a smearing region will lead to a breakdown
of the OPE, as the subleading twist terms will begin to dominate
the octet contribution. More quantitatively, we may expand the octet
contribution into an infinite sum of leading twist matrix
elements
\beqa 
\label{Bdcy}
Im T_8^\prime(^3P_J)=C_{^3P_J} g_s^2 \sum_{n=0}^\infty\frac{(-1)^n}{n !} \langle \Upsilon \mid \left[ \psi^\dagger \sigma_i \frac{i}{2}
 \stackrel{\leftrightarrow}{D}_j T^A 
\chi \right] iD_{\mu_1}...iD_{\mu_n}&&\!\!\!
\left[\chi^\dagger  \sigma_i \frac{i}{2}
 \stackrel{\leftrightarrow}{D}_j T^A 
\psi \right]\mid \Upsilon \rangle 
\\ \nonumber &&n^{\mu_1}...n^{\mu_n} \delta^{(n)}
(
2E_\gamma-2m_b),
\eeqa
\beqa 
\label{B1dcy}
Im T_8^\prime(^1S_0)=C_{^1S_0} g_s^2 \sum_{n=0}^\infty\frac{(-1)^n}{n !} 
\langle \Upsilon \mid \left[ \psi^\dagger   T^A 
\chi \right] iD_{\mu_1}...iD_{\mu_n}&&\!\!\!
\left[\chi^\dagger   T^A 
\psi \right]\mid \Upsilon \rangle 
\\ \nonumber &&n^{\mu_1}...n^{\mu_n} \delta^{(n)}
(
2E_\gamma-2m_b),
\eeqa
where $\delta^{(n)}$  denotes  $n$ derivatives
with respect to $2E_\gamma $  acting on the delta function. 
Eqs.(16,17) are similar to the expression found in the analysis of
the leptonic end-point spectrum in $B$ decays \cite{neub,sterman}.
An operator with $n$ factors of the covariant derivative 
scales as \cite{bbl} $v^{(6,7)+2n}$, as each derivative contributes 
a factor of $v^2$ to the scaling. If we smear, for example, using a Gaussian
of width $\Delta E_{\gamma}$ then for,  $\Delta E_{\gamma}<<m_bv^2$, the
terms in the sum grow with n, and the subleading-twist terms which
have been dropped cannot be neglected. On the other
hand, for $\Delta E_{\gamma}>>m_bv^2$,
the singlet contribution will dominate over the octet.
Thus, the finest energy resolution
with which we can examine the photon endpoint spectrum, 
without introducing yet more higher twist
structure functions, is of order
$m_bv^2$. Note that this corresponds to smearing over a range
of hadronic masses $\Delta m_X \sim m_b v$ which for large
$m_b$ is much greater than the QCD scale.
For the smearing width $\Delta E_{\gamma}\sim m_bv^2$
 the octet $^3 P_J$ contribution is of the same order
as the color-singlet since the naive octet suppression of $v^2$
is compensated by the fact that the singlet contribution starts
with a theta function, whereas the octet starts with a delta
function. The formally leading contribution \footnote{
See erratum in ref. \cite{bbl}.} however, comes from the octet
$^1S_0$ contribution which is enhanced by a power of $1/v$.
 Therefore, if we wish to resolve the photon energy spectrum within
$m_bv^2\simeq 500~MeV$ of the end-point, we must know the value
of the structure functions $f(k^+_8)_{^1S_0}$ and $f(k^+_8)_{^3P_J}$, 
which can be perhaps measured
by the lattice or extracted from another process.

Let us now consider the size of other possible large effects near the
end-point.
As mentioned in the introduction there is another nonperturbative
effect near the end point having to do with the hadronization
of the partons. The gluonic partons are not really massless, but
are actually finite invariant mass objects. This effect was 
addressed in ref. \cite{field}, where the author used a parton-shower
Monte Carlo approximation to determine the effects of hadronization on the
photon spectrum of the color-singlet contribution.
Hadronization cuts off the photon distribution at
the end point. This effect is important for hadronic masses, $m_X$,
 of order the QCD scale and should be subdominant to those of the octet
contribution. The success of hadronization models in describing the endpoint
spectrum may indicate that the color octet matrix elements are smaller than
expected on the basis of dimensional analysis.

In the end-point region 
there are also large perturbative corrections stemming from terms
of the form ln$(1-x)$ where $x=E_{\gamma}/m_b$. These terms result from the
fact that near the end-point gluon radiation is suppressed ruining the
delicate cancelation between real and virtual graphs. A resummation 
of these logarithms is expected to suppress the singlet spectrum near the end point.
The calculations in ref. \cite{phot} indicate that this resummation 
does indeed suppress the color-singlet rate near the end-point and are imporant
when $E_{\gamma} > 0.4 m_{\Upsilon}$. There will be similar
large logarithms in the Wilson coefficient for the octet contribution discussed
above, resumming these logarithms
should again suppress the end point region. A calculation of these effects
is necessary to undertake any phenomenological investigation of the
end-point region.

The results of this paper are applicable to both $c\bar{c}$ and
$b\bar{b}$ quarkonia states. However, we feel that the charm quark mass
is probably not large enough for our methods to be trustworthy.
We also note that effects similar to those discussed here are 
important for fragmentation into quarkonia near $z=1$ as well
as direct production in some kinematic circumstances\cite{us}. 
\vskip.2in
\noindent{\bf Acknowledgements}

While we this paper was being written we became aware of \cite{man}
which deals with issues similar to those here for the case of
$\eta_Q$ decay. I.Z.R is supported by  grant no. DOE-FG03-90ER40546.
M.B.W. is supported by grant no. DE-FG03-92-ER40701.

\end{document}